\documentclass{emulateapj}

\shorttitle{Detection of two periods in V2574 Oph}

\shortauthors{Kang et al.}

\begin{document}

\vskip 4 cm

\title {Detection of orbital and superhump periods in Nova V2574 Ophiuchi (2004)}
 
\author{Tae W. Kang$^{1}$, Alon Retter$^{1}$, Alex Liu$^{2}$, and Mercedes Richards$^{1}$}

\vskip 0.1 cm

\affil{$^{1}$Dept. of Astronomy \& Astrophysics, Penn State University, 525 Davey Lab, University Park, PA 16802; tkang@astro.psu.edu; retter@astro.psu.edu; mtr@astro.psu.edu}
\vskip 0.1 cm
\affil{$^{2}$Norcape Observatory, PO Box 300, Exmouth, 6707, Australia; asliu@onaustralia.com.au}
\vskip 0.1 cm

\begin{abstract}

We present the results of 37 nights of CCD unfiltered photometry of nova V2574 Oph (2004) from 2004 and 2005. We find two periods of 0.14164 d $\approx$ 3.40 h and 0.14773 d $\approx$ 3.55 h in the 2005 data. The 2004 data show variability on a similar timescale, but no coherent periodicity was found.  We suggest that the longer periodicity is the orbital period of the underlying binary system and that the shorter period represents a negative superhump. The 3.40 h period is about 4$\%$ shorter than the orbital period and obeys the relation between superhump period deficit and binary period. The detection of superhumps in the light curve is evidence of the presence of a precessing accretion disk in this binary system shortly after the nova outburst. From the maximum magnitude -- rate of decline relation, we estimate the decay rate $t_2$ = 17 $\pm$ 4 d and a maximum absolute visual magnitude of $M_{Vmax}$ = --7.7 $\pm$ 1.7 mag. 
\end{abstract}


\section{Introduction}

Novae are a subclass of cataclysmic variables (CVs) which contain a white dwarf and a solar-like companion that fills its Roche Lobe. Typically, the white dwarf accretes mass from an accretion disk that surrounds it. After of order ten thousand years, the white dwarf accumulates enough material for a thermonuclear-runaway event, which results in the observed nova outburst (e.g., Warner 1995). It is believed that the nova eruption disrupts the accretion disk, which is reformed several weeks or months later (e.g., Retter 2004).

Nova V2574 Oph ($\alpha_{2000.0}=17^h38^m45^s, \delta_{2000.0}= -23^o28'18''.5$) was discovered by Akira Takao (Kitakyushu, Japan) at V $\approx$ 10.2 on 2004 Apr. 14.80 UT (Yamaoka 2004). Mason et al. (2004) reported that echelle spectra (390 -- 900nm; resolution 48000) of V2574 Oph, obtained on 2004 Apr. 17.32 and 18.37 UT at La Silla with the 2.2-m telescope, were dominated by H$\alpha$, Na D, and Ca II emission lines, which are flanked by double P-Cyg profiles. From the minima of the P-Cyg absorptions, the measured expansion velocities were estimated as 400 and 1050 km s$^{-1}$ for H$\alpha$ and 400 and 1000 km s$^{-1}$ for Na D.

Rudy et al. (2004) observed V2574 Oph 73 days after maximum and found that the nova was still in an `Oxygen I' phase where the O I lines at 0.8446 and 1.1287 microns that are fluorescently excited by Lyman $\beta$ have strengths comparable to H$\alpha$. The optical region showed numerous multiplets of Fe II which characterize this class as novae. The low expansion velocities of the ejecta mentioned above seem to rule out the possibility that V2574 Oph is a hybrid nova, which evolved to the He/N class from the Fe II type (Della Valle \& Livio 1998). The overall appearance of the spectra indicated a substantial evolution with respect to early reports and suggested that this is a slow nova caught at maximum light (Bond \& Walter 2004). 

So far, there are about 50 novae with known orbital periods (Warner 2002). Typical nova periods range from about 2 to 9 h. Finding the orbital period of a nova yields an estimate of the secondary mass (e.g., Smith \& Dhillon 1998). In addition, detecting several periodicities in novae can help in classifying the system into different groups of CVs such as magnetic systems, intermediate polars, and/or permanent superhump systems (e.g., Diaz \& Steiner 1989; Baptista et al. 1993; Retter, Leibowitz \& Ofek 1997; Patterson et al. 1997; Skillman et al. 1997; Patterson \& Warner 1998; Retter \& Leibowitz 1998; Retter, Leibowitz \& Kovo-Kariti 1998; Patterson 1999; Retter, Leibowitz \& Naylor 1999; Skillman et al 1999; Retter \& Naylor 2000; Patterson 2001; Woudt \& Warner 2001; Lipkin et al. 2001; Patterson et al. 2002; Warner 2002; Woudt \& Warner 2002; Retter et al. 2003; Ak, Retter \& Liu 2005, Balman et al. 2005; see also Retter et al. 2002; Retter, Richards \& Wu 2005). This yields valuable information on the magnetic field of the white dwarf and reveals the presence or absence of the accretion disk. 

We have an ongoing program to observe novae with small telescopes to search for periodicities in their optical light curves. In this paper, we present extensive photometric observation of V2574 Oph, which suggests the presence of orbital and superhump periods.

\begin{table}\caption{The observations time table}
\begin{center}
\begin{tabular}{ccccc}

UT       & Time of Start    & Run Time  & Points  & Comments      	\\
(yymmdd) & (HJD --2453000)  &  (Hours)  & number                        \\
                                                                        \\
040531   & 157.07290        & 7.8       &   219     	&  	        \\
040622   & 179.00080        & 8.5       &   235	        & 		\\
040623   & 180.03790        & 7.4       &   205		& 		\\
040804   & 222.00030        & 5.2       &   386		& 		\\
040805   & 223.01050        & 5.1       &   371		& 		\\
040807   & 225.00060        & 5.2       &   286		& 		\\
040808   & 226.03010        & 4.0       &   177		&		\\
040809   & 227.01290        & 5.0       &   219		& 		\\
040810   & 227.99230        & 4.9       &   221		& 		\\
040812   & 229.98880        & 5.3       &   181		& 		\\
040813   & 231.00440        & 4.9       &   199		& 		\\
040814   & 232.01670        & 5.0       &   203		& 		\\
040815   & 233.01580        & 4.9       &   199		& 		\\
040816   & 234.00970        & 5.0       &   206		& 		\\
040817   & 235.01410        & 4.6       &   189		& 		\\ 
040818   & 236.03860        & 2.8       &   127	        & 		\\ 
040820   & 238.00240        & 4.5       &   191		& 		\\ 
040822   & 240.06770        & 2.9       &   129		& 		\\ 
040826   & 244.07070        & 3.1       &   142		& 		\\
040827   & 244.98720        & 4.2       &   190		& 		\\
040828   & 245.96630        & 4.7       &   209		& 		\\
040829   & 247.01970        & 3.8       &   161		& 		\\
040830   & 248.01820        & 3.9       &   152	        & 	        \\     
050507   & 498.14124 	    & 6.3       &   176		& 		\\
050508   & 499.14860        & 6.2       &   166		& 		\\
050509   & 500.14133  	    & 6.4       &   177		& 		\\
050510   & 501.14419  	    & 6.1       &   170		& 		\\
050511   & 502.13795  	    & 6.4       &   176		& 		\\
050512   & 503.14005  	    & 6.4       &   175		& 		\\
050513   & 504.14247  	    & 6.5       &   176		& 		\\
050514   & 505.14854  	    & 6.3       &   169		& 		\\
050516   & 507.13301  	    & 6.3       &   177		& 		\\
050607   & 529.08677        & 7.1       &   176         & clouds	\\ 
050610   & 532.08986        & 6.9       &   165         & clouds	\\
050611   & 533.08452        & 4.9       &   109 	& clouds	\\
050613   & 535.09411        & 6.6       &   187    	& 		\\
050614   & 536.06776        & 7.9       &   223    	& 		\\
\end{tabular}
\end{center}
\end{table}

\section{Observations}

V2574 Oph was observed on 23 nights during May, June, and August, 2004 and 14 nights during May and June, 2005. The observations span 379 nights and consist of 37 nights (203 hours in total). Table 1 presents a summary of the schedule of the observation. The photometry was carried out with a 0.3-m f/6.3 telescope coupled to an SBIG ST7E CCD camera. The pixel size of this CCD was 9 x 9 microns. This camera is attached to an Optec f5 focal reducer giving an image field of view of 15 x 10 arcmin. The range of seeing for the data was 2.5 to 3 arcsec. The telescope is located in Exmouth, Western Australia, and no filter was used. The exposure times were between 30 and 60 sec every 120 sec. The total number of our data points was 7219. Aperture photometry was used in the reduction, with an aperture size of 12 pixels. For the data taken during May and June in 2004, we estimated differential magnitudes with respect to GSC (Guide Star Catalog) 6827-640 (V = 12.6, the comparison star, denoted ``C''), using another star (1.2' S of the comparison star) in the field, which is not listed in the GSC, as the check star, ``K'' (V = 13.4). The magnitudes of the comparison stars were added to the differential magnitudes to give a rough estimate of the V magnitudes. With the fading of the nova and because of some instrumental constraints, different comparison stars were used for two other sub-groups of the data.  For 2004 August 4--10, we observed the following objects: a star, 0.76' SW of the variable as the comparison star and another star, 1.4' SW of the nova as the check star. Their estimated V magnitudes were 13.66 and 14.06, respectively. For the remaining nights, we obtained differential magnitudes with respect to a star, 3.42' SW of the variable, as the comparison star and another star, 3.1' S of the variable, as the check star. Their estimated V magnitudes were 12.30 and 13.40, respectively. The standard deviation of the errors in the V magnitudes was about 0.02 mag. The magnitudes of the stars were derived from the SBIG CCDOPS software.
 
Figure 1 displays the visual light curve of the nova from outburst until June 2005. The data were compiled from the Association Francaise des Observateurs d'Etoiles Variables (AFOEV) and from the American Association of Variable Star Observers (AAVSO). By combining the data from these associations of amateur observers with our data, we obtained 7308 individual points. The times of our observations are marked on the graph as well. Since our data were taken with an unfiltered CCD camera, we compared them with the AAVSO data when the observations were nearly simultaneous, and added 2 magnitudes to our estimates to compensate for the difference between the visual and unfiltered data. The light curve shows that the fading of the nova was relatively smooth. We estimate that the maximum visual magnitude was V = 9.5 $\pm$ 0.2.

We present the light curves of our observations of V2574 Oph obtained in 2004 and 2005 separately in Figure 2. The upper panel displays the  2004 light curve. We did not include the first three nights in this graph because of the large time gap and amplitude difference between these and the other nights. During the time interval spanned by these observations in 2004, the nova declined by 0.32 mag. The lower panel in Figure 2 represents the observations in 2005.

\begin{figure}
\epsscale{1} 
\plotone{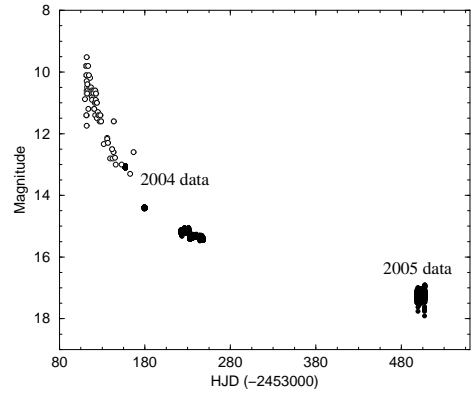}
\caption{The long term light curve of V2574 Oph. Empty circles represent visual estimates made by amateur astronomers, compiled by AFOEV and AAVSO. The filled circles correspond to our observations. Our data points were shifted by 2 magnitudes to compensate for the difference between visual and unfiltered data.}
\vskip 0.3 cm
\end{figure}

\section{Data Analysis}
\subsection{The long-term light curve of V2574 Oph}

By using the maximum visual magnitude of V = 9.5 $\pm$ 0.2 we measured the time required for a decline of two magnitudes from maximum $t_2$ = 17 $\pm$ 4 d. This makes V2574 Oph a fast nova according to the classification given in Table 5.4 of Warner (1995). This is inconsistent with the suggestion of Bond \& Walter (2004) that V2574 Oph is a slow nova (Section 1). We calculated the visual absolute magnitude at maximum brightness using Eq. 5.3 of Warner (1995), $M_{Vmax}$ = $a_2logt_2 + b_2$ where $a_2$ = 2.41 $\pm$ 0.23, $b_2$ = --10.70 $\pm$ 0.30 and $t_2$ = 17 $\pm$ 4 d, and obtained $M_{Vmax}$ = --7.7 $\pm$ 1.7 mag.

\subsection{The 2005 light curve}

We begin with our 2005 data since the periods were identified during this season. The noise level of the power spectrum is high when we combined the first run (first 9 nights) and the second run (last 5 nights) (Fig. 2b) because there was a 25-d gap between two runs. In addition, most nights in the second run in 2005 were affected by clouds. Thus, we decided to focus our analysis on the first 9 nights because they are almost consecutive nights and there is no large amplitude difference between the nightly means. A sample light curve from May 9, 2005 is shown in Figure 3. It shows a variation on a time scale of the order of 3 hours. 

The power spectrum (Scargle 1982) of the raw data of the first 9 nights in 2005 is displayed in Figure 4a. The power spectrum at mid-frequencies is dominated by two similar alias patterns around two central frequencies, $f_1$ = 7.060 d$^{-1}$ and $f_2$ = 6.769 d$^{-1}$, which correspond to the periodicities 0.14164 $\pm$ 0.00010 d and 0.14773 $\pm$ 0.00010 d. The highest peaks marked 1, 2 and 3 d$^{-1}$ are a result of the daily spacing between the nights. Those peaks at low frequencies are not real because the power spectrum of the check star minus comparison star (K-C) data has a similar pattern (see Fig. 4b). 

The power spectrum after subtracting the nightly trend from each night is displayed in Figure 4c. The 1, 2, and 3 d$^{-1}$ peaks disappeared. Figure 5a displays an expanded version of Figure 4a. It shows the two strong peaks, $f_1$ and $f_2$ with their $\pm$ 1 d$^{-1}$ aliases. Figure 5b represents the zoomed K-C power spectrum. 

\begin{figure}
\epsscale{1} 
\plotone{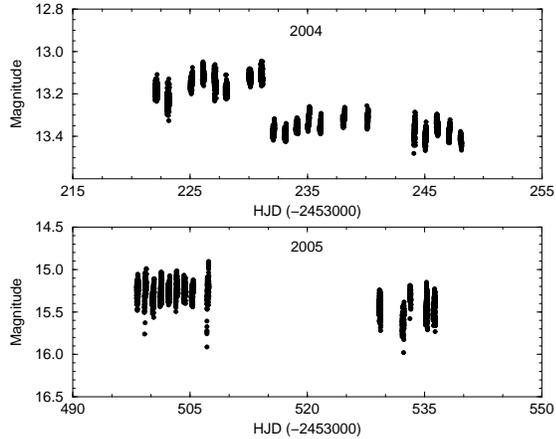}
\caption{The light curves of V2574 Oph obtained in the observatons in 2004 (upper panel) and 2005 (lower panel). The first three nights in 2004 are not shown because there was a large time gap and amplitude difference between these and the other nights.}
\vskip 0.3 cm
\end{figure} 

\subsubsection{Tests}

To confirm that the $f_1$ and $f_2$ peaks were real, we performed several tests. First, we checked to see if the $f_1$ peak in the power spectrum was an artefact of the window function. This was done by creating a noiseless simulation (i.e., planting a pure sinusoidal variation in the data with no errors) of the first 9 nights in the 2005 data. There was no evidence for significant power at the location of the peak.

We also checked whether the K-C power spectrum had a similar power and pattern near the $f_1$ peak. However, the K-C power spectrum did not show any strong power near the $f_1$ peak (see Figs. 4b and 5b).

To further check the significance of the $f_1$ peak, we used the bootstrap method. First, we scrambled the magnitude values, arbitrarily assigned them to the times of the observations, and calculated the corresponding power spectrum. Then, we found the power of the highest peak in the power spectrum. Finally, we plotted the histogram of the highest peaks of the 1000 simulations. This suggested that the $f_1$ peak was about 43$\sigma$ significant. Thus, we confirmed that the peak was real. The results were similar for the tests applied to the $f_2$ peak. 

\begin{figure}
\epsscale{1} 
\plotone{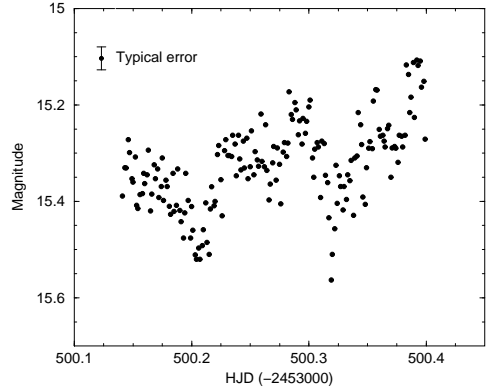}
\caption{A sample of a nightly light curve of V2574 Oph. The data were obtained May 9, 2005.}
\vskip 0.3 cm
\end{figure}

\subsubsection{Significance of the two peaks}

To check the significance of two independent periodicities in the light curve of the nova, we applied a few tests. First, we fitted and subtracted the first harmonic of the $f_2$ frequency from the data. The power spectrum of the residuals clearly showed $f_1$ as the dominant peak in the graph. Conversely, when the $f_1$ frequency was removed from the data, $f_2$ and its daily aliases dominated the residual power spectrum. We present these power spectra in Figure 4d and 4e, respectively. 

We also checked the significance of a second periodicity in the presence of the other one. We created an artificial light curve from the times of the actual observations, by superimposing a sinusoidal wave on one of the two periodicities over a random distribution of points representing white noise. The power spectrum of each of these synthetic light curves showed only the corresponding imposed periodicity, surrounded by an alias pattern similar to that of the actual data, with no trace of the other periodicity.

To check whether uncorrelated noise could be responsible for the presence of the candidates periods, we added noise to the model light curves of the first 9 nights in 2005. The noise in the original data was defined as the root mean square of the data minus the $f_1$ frequency. We then searched for the highest peak in a small interval (6.72--6.82 d$^{-1}$) around the $f_2$ peak. In 1000 simulations, no peak reached the height of the $f_2$ frequency. Similarly, for the $f_1$ frequency, we did 1000 simulations in the interval 7--7.1 d$^{-1}$. No peak reached the height of the $f_1$ peak either. So both periods seem to be real. 

As a final test for the presence of the two frequencies, we divided our 2005 data into two parts. In the power spectra of both parts of the observations, the same peaks appeared as the strongest ones in a 1 d$^{-1}$ frequency interval on both sides of the peaks (i.e., up to the 1 d$^{-1}$ aliases). Thus, we concluded that the two peaks in the 2005 data indicate real periodicities.  

In a search for a third frequency, we fitted and subtracted the $f_1$ and $f_2$ frequencies from the data. The power spectrum of the residuals did not show any addional significant peaks.  

\begin{figure}
\epsscale{1}
\plotone{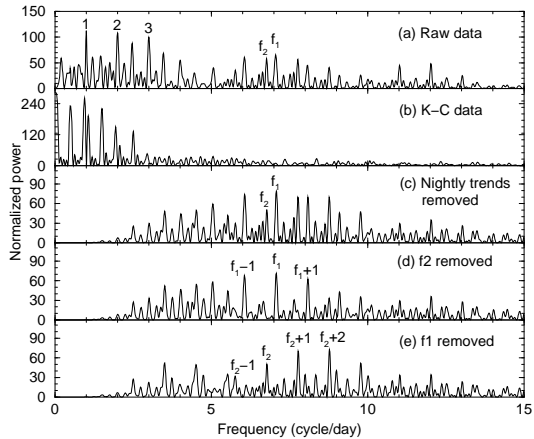}
\caption{Normalized power spectra of V2574 Oph in the first 9 nights in 2005. (a) The power spectrum of the raw light curve. (b) The power spectrum of the K-C data. (c) The power spectrum after subtracting the nightly trend from each night. The false peaks at low frequencies disappeared. (d) The power spectrum after removing the $f_2$ peak. It shows the $f_1$ peak with its $\pm$ 1 d$^{-1}$ aliases. (e) The power spectrum after removing the $f_1$ peak. It displays the $f_2$ peak with its $\pm$ 1 and 2 d$^{-1}$ aliases.}
\vskip 0.3 cm
\end{figure}

\subsection{The 2004 light curve}

The 2004 data were analyzed to search for the two periods found in the 2005 data. The 2004 data seem to have variations on timescales similar to the periodicities detected in the 2005 data. However, we could not find any convincing evidence of any coherent periodicity. This result is further discussed below.

\subsection{The structure of the periodicities}

In Figure 6 we present the first 9 nights in the 2005 data of V2574 Oph folded on the 0.14773 d period (top panel) and on the 0.14164 d period (bottom panel). The points in the figure are the average magnitude values in each of the 40 equal bins that cover the 0--1 phase interval. The peak to peak amplitudes of the mean variations were found to be 0.082 $\pm$ 0.007 and 0.077 $\pm$ 0.007 mag for the $f_1$ and $f_2$ frequencies, respectively. The amplitudes were derived by fitting a sinusoidal function to the mean light curve.

The best fitted ephemerides of the periodicities are: 
\vskip 0.2 cm
$T_{1(min)}$(HJD) = 2453498.215 + 0.14164E 

\hskip96pt $\pm$0.040 $\pm$ 0.00010                 

\vskip 0.2cm
$T_{2(min)}$(HJD) = 2453498.117 + 0.14773E

\hskip96pt $\pm$0.041 $\pm$ 0.00010
         
\begin{figure}
\epsscale{1}
\plotone{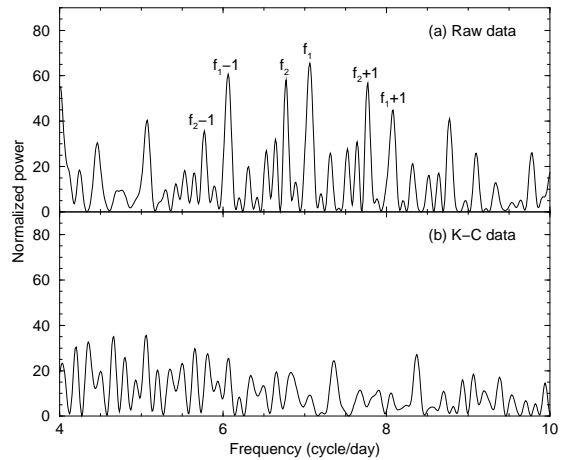}
\caption{(a) The power spectrum of the first 9 nights in 2005 (Fig. 4a) zoomed into the 4--10 d$^{-1}$ range of frequencies. The two frequencies ($f_1$ and $f_2$) are marked as well as their $\pm$ 1 d$^{-1}$ aliases.(b) The power spectrum of the K-C data. It doesn not display any significant periodicity.}
\vskip 0.3 cm
\end{figure} 

\section{Discussion}

The photometric data of V2574 Oph show the presence of two independent periodicities in the light curve: $P_1$ = 3.40 h and $P_2$ = 3.55 h, which correspond to $f_1$ and $f_2$, respectively. The orbital period distribution of novae has a peak at 3--4 h (Diaz \& Bruch 1997; Warner 2002). Thus, we suggest that one of the periods we found is the orbital period of the underlying binary system. We explore two possibilities, namely that the second periodicity could result from a spin period in a nearly synchronous polar system or a superhump.
 
\subsection{A nearly synchronous polar?}

 The AM Her stars (magnetic CVs) differ from the non-magnetic systems in two important qualitative respects: (1) a strong magnetic field on the primary star funnels the infalling gas onto one or two localized accretion shocks near the white dwarf's magnetic pole(s), where X-ray Bremsstrahlung and polarized optical/infrared cyclotron emission arise; and (2) the white dwarf spin and binary orbital motions are locked in a rigid corotating geometry (e.g., Schmidt \& Stockman 1991).

Nearly synchronous polars are a subclass of magnetic CVs, sharing many of their properties with AM Her stars, but having a white dwarf that rotates with a period that differs by $\sim$1$\%$ from the orbital period. There are four known nearly synchronous polars: V1500 Cyg (e.g., Schmidt, Liebert \& Stockman 1995), BY Cam (e.g., Mason et al. 1998), V1432 Aql (e.g., Geckeler \& Staubert 1997), RX J2115-5840 (Ramsay et al. 2000), and one candidate -- V4633 Sgr (Lipkin et al. 2001).

Warner (1995) explained that as the primary of the binary system rotates asynchronously, the accretion flow is more variable than for a phase-locked polar. Accretion occurs on both magnetic poles, but at any given time accretion onto the pole nearest the secondary is most favored.  

The reason for the asynchronization was proposed as a recent nova outburst (Stockman, Schmidt \& Lamb 1988). In one case, V1500 Cyg, the asynchronous rotation is clearly associated with its nova eruption in 1975. Two other nearly synchronous polars were suggested to have also undergone a recent nova event: V1432~Aql (Schmidt \& Stockman 2001) and BY Cam (Bonnet-Bidaud \& Mouchet 1987). Similarly, V4633 Sgr, which was proposed as a nearly synchronous system (Lipkin et al. 2001), is a post-nova. 

The difference between the two periods of V2574 Oph that we found in the 2005 data, about one year after outburst is $\sim$4$\%$. This is much larger than all values found in the nearly synchronous polars (Lipkin et al 2001). This difference can not be attributed to the proximity in time of our observations to the nova outburst because the low value of asynchonizatoin in V1500 Cyg was already observed about three months after its nova outburst (Semeniuk, Kruszewski \& Schwarzenberg-Czerny 1976). This is much sooner after the nova event than our observations. Thus, a nearly synchronous polar model does not seem to apply to V2574 Oph.

\begin{figure}
\epsscale{1} 
\plotone{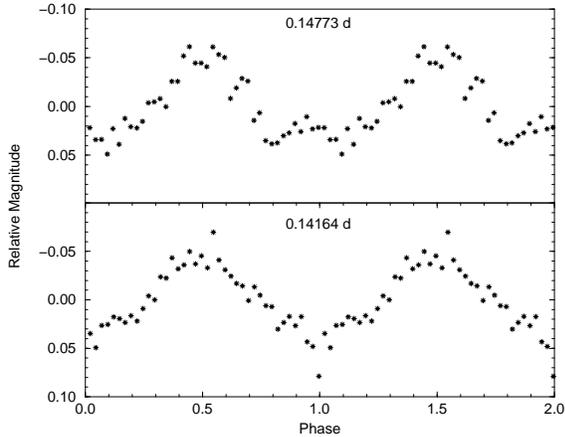}
\caption{The light curve of V2574 Oph obtained in the first nine nights in 2005, folded on the 0.14773 d period (top panel) and the 0.14164 d period (bottom panel) and binned into 40 equal bins. Two cycles are shown for clarity.}
\vskip 0.3 cm
\end{figure}

\subsection{Permanent superhumps}

Patterson \& Richman (1991) initially suggested the term `permanent superhump' for the subclass of CVs whose light curves show quasi-periodicities slightly different from their binary orbital periods. Unlike SU UMa systems (see Warner 1995 for a review of SU UMa systems and CVs in general), which display this behavior only during superoutbursts, permanent superhump systems show the phenomenon during their normal brightness state. 

Whitehurst \& King (1991) suggested that superhumps occur when the accretion disk extends beyond the 3:1 resonance radius. According to Osaki (1996), permanent superhump systems differ from other subclasses of non-magnetic CVs by having relatively short orbital periods and high mass-transfer rates, resulting in accretion disks that are thermally stable but tidally unstable. Retter \& Naylor (2000) provided observational support for this idea.

In permanent superhump systems, periods longer and shorter than the orbital periods have been observed. These are called positive and negative superhumps respectively. A positive superhump, a periodicity that is a few per cent larger than the orbital period, is explained by the beat period between the binary motion and the precession of an eccentric accretion disk in the apsidal plane. Observations of positive superhumps showed a roughly linear relation between the period excess, expressed as a fraction of the binary period and the binary period itself (Stolz \& Schoembs 1984; Retter et al. 1997; Patterson 1999). A negative superhump is a period slightly shorter than the orbital period. It is explained by the beat periodicity between the nodal precession of the accretion disk and the orbital period (Patterson et al. 1993; Patterson 1999). Negative superhumps also show a roughly linear correlation between the period deficit and the binary period itself (Patterson 1999), and Retter et al. (2002) proposed that the ratio of the negative superhump deficit over the positive superhump excess in systems that show both types of superhumps is connected with the orbital period. 

Two periods were detected in the light curve of V2574 Oph from the 2005 data (Section 3.2). If $P_1$ = 3.40 h is the orbital period, then $P_2$ = 3.55 h is a positive superhump. Alternatively, if $P_2$ is the orbital period, then $P_1$ is a negative superhump. Using $\Delta$P $\approx$ 4$\%$, we can check which of the two scenarios better fits the two relations for positive and negative superhumps. In case $P_1$ is the orbital period, $P_2$ is a positive superhump, and the superhump period excess is then about 4$\%$. According to the relation shown in figure 1. in Patterson (1999), the positive superhump excess should be $\sim$7$\%$ for an orbital period of 3.40 h. Therefore, this case does not seem to apply to V2574 Oph. If $P_2$ is the orbital period, $P_1$ is a negative superhump, and the superhump period deficit is then about 4$\%$. This nicely fits the relation in figure 1 of Patterson (1999). Thus, we interpret $P_2$ = 3.55 h as the orbital period and $P_1$ = 3.40 h as the negative superhump period.    

\subsection{The presence of the accretion disk}

It was believed that the accretion disk around the white dwarf is destroyed by the nova outburst and that it takes a few decades for the disk to reform. However, permanent superhumps have been observed in V1974 Cyg about two years after its nova outburst (Retter et al. 1997; Skillman et al. 1997; Retter 1999). The detection of superhumps is evidence of the early presence of the precessing accretion disk in this system. Similarly, the presence of superhumps in V2574 Oph suggests that an elliptic accretion disk existed in V2574 Oph about one year after outburst.

\subsection{The 2004 data}
 
We did not detect any coherent frequencies in the 2004 data (Section 3.3). We suspect that no periods were found in these data because of three reasons: (1) the light curve might bave been affected by a significant contribution from the nebula and/or by optically thick winds; (2) the accretion disk might have been in an unstable state at that time and could have still been reforming thus adding more noise and reducing the chances of detecting any frequencies; and (3) besides the first three nights, which were obtained very early after the nova eruption, the nightly runs in the 2004 data were shorter than in the 2005 data (see Table 1). Therefore, the results from the 2004 data were less reliable. 
 
\vskip 0.2 cm

\section{SUMMARY AND CONCLUSION}

(1) By using the maximum visual magnitude of V = 9.5 $\pm$ 0.2 we measured the decay time $t_2$ = 17 $\pm$ 4 d from the long-term light curve. This makes V2574 Oph a fast nova. We estimated a visual absolute magnitude in maximum of $M_{Vmax}$ = --7.7 $\pm$ 1.7 mag.

(2) We found two periods of 0.14164 d $\approx$ 3.40 h and 0.14773 d $\approx$ 3.55 h in the 2005 data. 

(3) We interpret the longer 3.55 h period as the orbital period of the binary system and the 3.40 h period as a negative superhump period.

(4) More observations (including radial velocity studies) are required to confirm our results and to follow the evolution of the periodicities in time. 

\section{acknowledgements}

We acknowledge the amateur observers of the AAVSO and AFOEV who made the observations that comprise the long-term light curve of nova V2574 Oph used in this study. TWK was supported by an REU supplement to NSF grant AST -- 0434234 (PI -- G. J. Babu). AR was partially supported by a research associate fellowship from Penn State University. We thank Jan Budaj for useful comments on an early version of the paper. We also thank the anonymous referee for very useful comments that significantly improved the paper. 


\bibliography{%
/home-astron/retter/bib/mnemonic,%
/home-astron/retter/bib/mnemonic-simple,%
/home-astron/retter/bib/all_algol%
}

\begin{thebibliography}{24}
\expandafter\ifx\csname natexlab\endcsname\relax\def\natexlab#1{#1}\fi
\expandafter\ifx\csname url\endcsname\relax
  \def\url#1{{\tt #1}}\fi

\bibitem[{}]{506}

\bibitem[]{255} Ak, T., Retter, A., \& Liu, A. 2005, PASA, 22, 298

\bibitem[]{258} Balman, S., Yilmaz, A., Retter, A., Saygac, T., \& Esenoglu, H. 2005, MNRAS, 356, 773

\bibitem{b2} Baptista, R., Yablonski, F.J., Cieslinski, D., \& Steiner, J.E.,
1993, ApJ, 406, L67


\bibitem[]{265} Bond, H. E., \& Walter, F. 2004, IAUC. 8324

\bibitem[]{268} Bonnet-Bidaud, J. M., \& Mouchet, M. 1987, A\&A, 188, 89

\bibitem[]{}Della Valle, M., \& Livio, M. 1998, ApJ, 506, 818

\bibitem[]{270} Diaz, M. P., \& Bruch, A. 1997, A\&A, 322, 807

\bibitem[]{275} Diaz, M.P., \& Steiner, J.E. 1989, ApJ, 339, L41

\bibitem[]{278} Geckeler, R. D., \& Staubert, R. 1997, A\&A, 325, 1070

\bibitem[]{280} Lipkin, Y., Leibowitz, E. M., Retter, A., \& Shemmer, O. 2001, MNRAS, 328, 1169.  

\bibitem[]{284} Mason, E., Ederoclite, A., Salas, F., \& Della Valle, M. 2004, IAUC., 8328

\bibitem[]{287} Mason, P. A., Ramsay, G., Andronov, I., Kolesnikov, S., Shakhovskoy, N., \& Ravlenko, E. 1998, MNRAS, 295, 51

\bibitem[]{289} Osaki, Y. 1996, PASP, 108, 39

\bibitem[]{292} Patterson, J. 1999, in Mineshige S., Wheeler C., eds, 
Disk Instabilities in Close Binary Systems, Universal Academy Press, 
Tokyo, p. 61

\bibitem[]{297} Patterson, J. 2001, PASP, 113, 736

\bibitem{b3} Patterson, J., Kemp, J., Saad, J., Shambrook, A., Thomas, G. R., 
Halpern, J., \& Skillman, D. 1997a, PASP, 109, 1100

\bibitem[]{304} Patterson, J., \& Richman, H. 1991, PASP, 103, 735

\bibitem[]{307} Patterson, J. Thomas, G. R., Skillman, D., Diaz, M., \& Suleimanov, V. F. 1993, ApJS, 86, 235

\bibitem[]{310} Patterson, J., \& Warner, B. 1998, PASP, 110, 1026

\bibitem[]{313} Patterson, J. et al. 2002, PASP, 114, 65

\bibitem[]{316} Ramsay, G., Potter, S., Cropper, M., Buckley, D. A. H., \& Harrop-Allin, M. K. 2000, MNRAS, 316, 225

\bibitem[]{319} Retter, A. 1999, PASP, 111, 774

\bibitem[]{322} Retter, A. 2004, ApJ, 615, 125

\bibitem[]{325} Retter, A., \& Leibowitz, E. M. 1998, MNRAS, 296, L37

\bibitem[]{328} Retter, A., Chou, Y., Bedding, T., \& Naylor, T. 2002, MNRAS, 330, L37

\bibitem[]{332} Retter, A., Hellier, C., Augusteijn, T., Naylor, T., Bedding, T.,
Bembrick, C., McCormick, J., \& Velthuis, F. 2003, MNRAS, 340, 679 

\bibitem[]{336} Retter, A., \& Naylor, T. 2000, MNRAS, 319, 510 

\bibitem[]{339} Retter, A., Leibowitz, E. M., \& Naylor, T. 1999, MNRAS, 308, 140

\bibitem[]{343} Retter, A., Leibowitz, E. M., \& Ofek, E. O. 1997, MNRAS, 286, 745

\bibitem{b12} Retter, A., Leibowitz, E. M., \& Kovo-Kariti, O. 1998, MNRAS, 293, 145

\bibitem[]{349} Retter, A., Richards, M. T., \& Wu, K. 2005, ApJ, 621, 417

\bibitem[]{352} Rudy, R. J., Lynch, D. K., Mazuk, S. M., Venturini, C. C., Puetter, R. C., \& Perry, R. B. 2004, AASM 205, 19.20

\bibitem[]{355} Scargle, J. D. 1982, ApJ, 263, 835

\bibitem[]{359} Schmidt, G. D., Liebert J., \& Stockman, H. S. 1995, ApJ, 441, 414

\bibitem[]{361} Schmidt, G. D., \& Stockman, H. S. 1991, ApJ, 371, 749

\bibitem[]{364} Schmidt, G. D., \& Stockman, H. S. 2001, ApJ, 548, 410

\bibitem{} Semeniuk, I., Kruszewski, A., \& Schwarzenberg-Czerny, A. 1976, IBVS, 1157, 1

\bibitem[]{367} Skillman, D. R., Harvey, D., Patterson, J., \& Vanmunster, T. 1997,
PASP, 109, 114

\bibitem[]{371} Skillman, D. R., Patterson, J., Kemp, J., Harvey, D. A., 
Fried, R., Retter, A., Lipkin, Y., \& Vanmunster, T. 1999, PASP, 111, 1281

\bibitem[]{375} Smith, D. A., \& Dhillon, V. S. 1998, MNRAS, 301, 767

\bibitem[]{378} Stockman, H. J., Schmidt, G. D., \& Lamb, D. Q. 1988, ApJ, 441, 414

\bibitem[]{380} Stolz, B., \& Schoembs, R. 1984, A\&A, 132, 187

\bibitem[]{383} Warner, B. 1995, Cataclysmic Variable Stars. Cambridge Univ. Press, Cambridge

\bibitem[]{385} Warner, B. 2002, in ``Classical Nova explosions'', eds. 
Hernanz M., Jos\'{e} J., AIP Conf. Ser., Vol 637, 3

\bibitem[]{389} Whitehurst, R., \& King, A. 1991 MNRAS, 249, 25

\bibitem[]{392} Woudt, P. A., \& Warner, B. 2001, MNRAS, 328, 159

\bibitem[]{396} Woudt, P. A., \& Warner, B. 2002, MNRAS, 335, 44

\bibitem[]{400} Yamaoka, H. 2004, IAUC., 8323

\end{thebibliography}

\bibliographystyle{/home-astron/bedding/bstinputs/natbib/mynatbib}
\end{document}